\documentclass{article}
\usepackage[margin=1in]{geometry}
\usepackage{amsmath, amssymb, amsthm}
\usepackage{graphicx}
\usepackage{hyperref}
\usepackage{xcolor}
\usepackage{natbib}

\hypersetup{
  colorlinks=true,
  linkcolor=blue,
  citecolor=blue,
  urlcolor=blue
}

\title{Mathematical Supplement for the \texttt{gsplat} Library}
\author{Vickie Ye\quad Angjoo Kanazawa\\UC Berkeley}
\date{}

\begin{document}

\maketitle

\section{Introduction}
This report provides the mathematical details of the \texttt{gsplat} library,
a modular toolbox for efficient differentiable Gaussian splatting, as proposed by \citet{kerbl20233d}.
We aim to provide a self-contained reference for the computations involved in the forward and backward passes of differentiable Gaussian splatting.
We first review background for projecting and rasterizing 3D Gaussians into an output image in Section~\ref{sec:forward}.
We then derive the gradients of a loss on the rendered images with respect to the 3D Gaussian parameters for the backward pass in Section~\ref{sec:backward}.
To facilitate practical usage and development,
we provide a user-friendly Python API that exposes each component of the forward and backward passes in rasterization
at \href{https://github.com/nerfstudio-project/gsplat}{github.com/nerfstudio-project/gsplat}.



\section{Rasterization Forward Pass}
\label{sec:forward}
A 3D Gaussian is parameterized by its mean $\mu \in \mathbb{R}^3$,
covariance $\Sigma \in \mathbb{R}^{3\times 3}$,
color $c \in \mathbb{R}^3$, and opacity $o \in \mathbb{R}$.
To render a view of the Gaussians, we first compute their projected 2D locations and extents in the camera plane.
The visible 2D Gaussians are then sorted by depth and composited from front to back to construct the output image.

\subsection{Projection of Gaussians}
The render camera is described by its extrinsics $T_{\textrm{cw}}$,
which transforms points from the world coordinate space to the camera coordinate space,
and its intrinsics, which are the focal length $(f_x, f_y)$ and the principal point $(c_x, c_y)$ of the camera plane.
We write the transformation from camera space to normalized clip space with the projection matrix $P$.
\begin{equation}
T_{\textrm{cw}}  = \begin{bmatrix}
    R_{\textrm{cw}} & t_{\textrm{cw}}\\
    0 & 1
\end{bmatrix} \in SE(3), \quad
P = \begin{bmatrix}
    2f_x / w & 0 & 0 & 0\\
    0 & 2f_y / h & 0 & 0 \\
    0 & 0 & (f+n)/(f-n) & -2fn / (f - n)\\
    0 & 0 & 1 & 0
\end{bmatrix},
\end{equation}
where $(w, h)$ are the output image width and height, and $(n, f)$ are the near and far clipping planes.
We project the 3D mean $\mu$ into pixel space via standard perspective projection.
We transform the mean $\mu$ into $t \in \mathbb{R}^4$ in camera coordinates,
$t' \in \mathbb{R}^4$ in ND coordinates, and 
$\mu' \in \mathbb{R}^2$ in pixel coordinates
\begin{equation}
    t = T_{\textrm{cw}} \begin{bmatrix} \mu & 1 \end{bmatrix}^\top,
    \quad t' = P t,
    \quad \mu' = \begin{bmatrix}
        (w \cdot t'_x / t'_w + 1) / 2 + c_x \\
        (h \cdot t'_y / t'_w + 1) / 2 + c_y \\
    \end{bmatrix},
\end{equation}
where $w$ and $h$ are the output image width and height respectively.

Perspective projection of a 3D Gaussian does not result in a 2D Gaussian.
We approximate the projection of $\Sigma$ to pixel space with a first-order Taylor expansion at $t$ in the camera frame.
Specifically, we compute the affine transform
$J \in \mathbb{R}^{2\times 3}$ as shown in \cite{zwicker2002ewa} as
\begin{equation}
    J = \begin{bmatrix}
    f_x / t_z & 0 & -f_x \cdot t_x / t_z^2 \\
    0 & f_y / t_z & -f_y \cdot t_y / t_z^2 \\
    \end{bmatrix}.
\end{equation}

The 2D covariance matrix $\Sigma' \in \mathbb{R}^{2\times 2}$ is then given by
\begin{equation}
    \Sigma' = J R_{\textrm{cw}} \Sigma R_{\textrm{cw}}^\top J^\top.
\end{equation}

Finally, we parameterize the 3D covariance $\Sigma$ with scale $s \in \mathbb{R}^3$
and rotation quaternion $ q \in \mathbb{R}^4$ and convert it to $\Sigma$.
We first convert the quaternion $q = (x, y, z, w)$ into a rotation matrix,
\begin{equation}
R =  \begin{bmatrix}
        1 - 2 \cdot (y^2 + z^2) & 2 \cdot (x y - w z) & 2 \cdot (x z + w y)\\
        2 \cdot (x y + w z) & 1 - 2 \cdot (x^2 - z^2) & 2 \cdot (y z - w x)\\
        2 \cdot (x z - w y) & 2 \cdot (y z + w x) & 1 - 2 \cdot (x^2 + y^2)
    \end{bmatrix}.
\end{equation}
The 3D covariance $\Sigma$ is then given by 
\begin{equation}
    \Sigma = RS S^\top R^\top,
\end{equation}
where $S = \mathrm{diag}(s) \in \mathbb{R}^{3\times 3}$.

\subsection{Depth Compositing of Gaussians}
We directly follow the tile sorting method introduced by~\citep{kerbl20233d}, 
which bins the 2D Gaussians into $16\times 16$ tiles and sorts them per tile by depth.
For each Gaussian, we compute the axis-aligned bounding box around the 99\% confidence ellipse of each 2D projected covariance (3 sigma),
and include it in a tile bin if its bounding box intersects with the tile.
We then apply the tile sorting algorithm as presented in Appendix C of~\citep{kerbl20233d}
to get a list of Gaussians sorted by depth for each tile.

We then rasterize the sorted Gaussians within each tile.
For a color at a pixel $i$, let $n$ index the $N$ Gaussians involved in that pixel.
\begin{equation}
    C_i = \sum_{n \le N} c_n \cdot \alpha_n \cdot T_n, ~~\textrm{where}~~ T_n = \prod_{m < n} (1 - \alpha_m).
\end{equation}
We compute $\alpha$ with the 2D covariance $\Sigma' \in \mathbb{R}^{2\times 2}$ and opacity parameters:
\begin{align*}
    \alpha_n = o_n \cdot \exp(-\sigma_n), \quad
    \sigma_n = \frac{1}{2} \Delta_n^\top \Sigma'^{-1} \Delta_n,
\end{align*}
where $\Delta \in \mathbb{R}^2$ and is the offset between the pixel center and the 2D Gaussian center $\mu' \in \mathbb{R}^2$.
We compute $T_n$ online as we iterate through the Gaussians front to back.

\newcommand{\grad}[2]{\frac{\partial{#1}}{\partial{#2}}}
\newcommand{\fprod}[1]{\langle #1 \rangle}
\newcommand{\cL}{\mathcal{L}}
\newcommand{\bR}{\mathbb{R}}

\section{Computing Gradients of Gaussians}
\label{sec:backward}


We now compute the gradients of a scalar loss with respect to the input Gaussian parameters.
That is, given the gradient of a scalar loss $\cL$ with respect each pixel of the output image, we propagate the gradients backward toward the original input parameters with standard chain rule mechanics.

In the following we will use the Frobenius inner product in deriving the matrix derivatives \citep{giles2008extended}:
\begin{equation}
    \fprod{X, Y} = \mathrm{Tr}(X^\top Y) = \mathrm{vec}(X)^\top \mathrm{vec}(Y) \in \mathbb{R},
\end{equation}
and can be thought of as a matrix dot product.
The Frobenius inner product has the following properties:
\begin{align}
    \fprod{X, Y} &= \fprod{Y, X},\\
    \fprod{X, Y} &= \fprod{X^\top, Y^\top}, \\ 
    \fprod{X, Y Z} &= \fprod{Y^\top X, Z} = \fprod{X Z^\top, Y}, \\
    \fprod{X, Y + Z} &= \fprod{X, Y} + \fprod{X, Z}.
\end{align}

Suppose we have a scalar function $f$ of $X \in \mathbb{R}^{m\times n}$, 
and that $X = A Y$, with $A \in \mathbb{R}^{m\times p}$ and $Y \in \mathbb{R}^{p \times n}$.
We can write the gradient of $f$ with respect to an arbitrary scalar $x \in \bR$ as
\begin{equation}
    \grad{f}{x}
    = \fprod{ \grad{f}{X}, \grad{X}{x} },
\end{equation}
for which we use the shorthand
\begin{equation}
\partial f = \fprod{\grad{f}{X}, \partial X}.
\end{equation}
Here, $\grad{f}{x} \in \mathbb{R}$,
$\grad{f}{X} \in \mathbb{R}^{m\times n}$,
and $\grad{X}{x} \in \mathbb{R}^{m\times n}$.

In this case, it is simple to continue the chain rule.
Letting $G = \grad{f}{X}$, we have
\begin{align*}
    \grad f x &= \fprod{ G, \grad{(AY)}{x} } \\
    &= \fprod{G, \grad{A}{x} Y} + \fprod{ G, A \grad{Y}{x} } \\
    &= \fprod{ G Y^\top, \grad{A}{x} } + \fprod{ A^\top G, \grad{Y}{x} }.
\end{align*}
From here,
we read out the elements of
the gradients of $f$ with respect to $A$ and $Y$
by letting $x = A_{ij}$ and $x = Y_{ij}$ respectively,
and find that
\begin{equation}
    \grad{f}{A} = G Y^\top \in \mathbb{R}^{m\times p}, \quad
    \grad{f}{Y} = A^\top G \in \mathbb{R}^{p\times n}.
\end{equation}

\subsection{Depth Compositing Gradients}

We start with propagating the loss gradients of a pixel $i$
back to the Gaussians that contributed to the pixel.
Specifically, for a Gaussian $n$ that contributes to the pixel $i$, 
we compute the gradients with respect to color $\grad{\cL}{c_n} \in \bR^3$,
opacity $\grad{\cL}{o_n} \in \bR$,
the 2D means $\grad{\cL}{\mu'_n} \in \bR^2$,
and 2D covariances $\grad{\cL}{\Sigma'_n} \in \bR^{2\times 2}$,
given the $\grad{\cL}{C_i} \in \bR^3$.
In the forward pass, we compute the contribution of each Gaussian to the pixel color from front to back,
i.e. Gaussians in the back are downstream of those in the front.
As such, in the backward pass, we compute the gradients of the Gaussians from back to front.


For the color, we have
\begin{equation}
\grad{C_i(k)}{c_n(k)} =  \alpha_n \cdot T_n
\end{equation}
for each channel $k$.
We save the final $T_N$ value from the forward pass and compute next $T_{n-1}$ values as we iterate backward:
\begin{equation}
    T_{n-1} = \frac{T_n}{1 - \alpha_{n-1}}.
\end{equation}

For the $\alpha$ gradient, for each channel $k$ we have the scalar gradients
\begin{equation}
    \grad{C_i(k)}{\alpha_n}
    = c_n(k) \cdot T_n - \frac{S_n(k)}{1-\alpha_n}
    ~~\mathrm{where}~~
    S_n = \sum_{m > n} c_m \alpha_m T_m.
\end{equation}
We can also compute $S_{n-1}$ as we iterate backward over Gaussians:
\begin{align*}
S_N(k) &= 0\\
S_{n-1}(k) &= c_n(k) \alpha_n T_n + S_n(k).
\end{align*}

For the opacity and sigma, we have scalar gradients
\begin{align}
    \grad{\alpha_n}{o_n}
    = \exp(-\sigma_n),
    \quad
    \grad{\alpha_n}{\sigma_n}
    = -o_n \exp(-\sigma_n).
\end{align}

For the 2D mean, we have the Jacobian
\begin{align*}
    \grad{\sigma_n}{\mu'_n} =  \grad{\sigma_n}{\Delta_n}
    = \Sigma'^{-1}_n \Delta_n \in \mathbb{R}^2.
\end{align*}

For the 2D covariance, we let $Y = \Sigma'^{-1}_n$, which has a straightforward Jacobian from $\sigma_n$:
\begin{align*}
    \grad{\sigma_n}{Y}
    &= \frac{1}{2} \Delta_n \Delta_n^\top \in \bR^{2\times 2}.
\end{align*}


To continue back-propagating through $Y \in \bR^{2\times 2}$,
we let $G = \grad{\sigma_n}{Y}$
and write the gradients with respect to a scalar variable $x$ as 
\begin{equation}
\grad{\sigma_n}{x} = \fprod{G, \grad{Y}{x}}.
\end{equation}

We use the identity \citep{petersen2008matrix, dwyer1948} that
$\grad{Y}{x} = -Y \grad{\Sigma'_n}{x} Y$,
and have
\begin{align*}
    \grad{\sigma_n}{x}
    &= \fprod{G, -Y \grad{\Sigma'_n}{x} Y} \\
    &= \fprod{-Y^\top G Y^\top, \grad{\Sigma'_n}{x} }
\end{align*}
The gradient of $\sigma_n$ with respect to $\Sigma'_n$ is then
\begin{equation}
\grad{\sigma_n}{\Sigma'_n}
= -\frac{1}{2} \Sigma'^{-1}_n \Delta_n \Delta_n^\top \Sigma'^{-1}_n.
\end{equation}

\subsection{Projection Gradients}
Given the gradients of $\cL$ with respect the projected 2D mean $\mu'$ and covariance $\Sigma'$ of a Gaussian,
we can continue to backpropagate the gradients of its 3D means $\mu$ and covariances $\Sigma$.
Here we deal only with a single Gaussian at a time, so we drop the subscript $n$
and compute the gradients $\grad{\cL}{\mu} \in \bR^3$ and $\grad{\cL}{\Sigma} \in \bR^{3\times 3}$,
given the gradients $\grad{\cL}{\mu'} \in \bR^2$ and $\grad{\cL}{\Sigma'} \in \bR^{2\times 2}$.

We first compute the gradient contribution of 2D mean $\mu'$ to camera coordinates $t \in \bR^4$,
and of 2D covariance $\Sigma'$ to 3D covariance $\Sigma$ and camera coordinates $t$.
Note that both $\mu'$ and $\Sigma'$ contribute to the gradient with respect to $t$:
\begin{equation}
\grad{\cL}{t_i}
= \grad{\cL_{\mu'}}{t_i} + \grad{\cL_{\Sigma'}}{t_i}
= \grad{\cL}{\mu'} \grad{\mu'}{t_i} + \fprod{\grad{\cL}{\Sigma'}, \grad{\Sigma'}{t_i}}
\label{eq:dLdt}
\end{equation}

For 2D mean $\mu'$, we have the contribution to the gradient of $t$ as
\begin{equation}
\grad{\cL_{\mu'}}{t} = 
\frac{1}{2} P^\top \begin{bmatrix}
    w / t_w & 0 & 0 & -w\cdot t_x/ t_w^2\\
    0 & h / t_w & 0 & -w\cdot t_y/ t_w^2
\end{bmatrix}^\top \grad{\cL}{\mu'}.
\end{equation}

The 2D covariance $\Sigma'$ contributes to the gradients of $\Sigma$ and $t$.
where $\Sigma' = T \Sigma T^\top$.
The contribution to $\Sigma$ is straightforward.
Letting $G = \grad{\cL}{\Sigma'}$, we have
\begin{align*}
    \partial \cL_{\Sigma'} &= \fprod{G, \partial \Sigma'}\\
    &= \fprod{G, (\partial T) \Sigma T^\top + T (\partial \Sigma) T^\top + T \Sigma (\partial T^\top) } \\
    &= \fprod{GT\Sigma^\top, \partial T} + 
    \fprod{T^\top G T, \partial \Sigma} + 
    \fprod{G^\top T \Sigma, \partial T} \\
    &= \fprod{GT\Sigma^\top + G^\top T \Sigma, \partial T} + 
    \fprod{T^\top G T, \partial \Sigma}.
\end{align*}
We read out the gradient with respect to $\Sigma \in \bR^{3\times 3}$ as
\begin{equation}
\grad{\cL}{\Sigma}
= T^\top \grad{\cL}{\Sigma'} T.
\end{equation}

We continue to propagate gradients through $T = J R_\textrm{cw} \in \bR^{2\times 3}$ for $J \in \bR^{2\times 3}$:
\begin{equation}
\partial \cL
    = \fprod{\grad{\cL}{T}, (\partial J) R_\textrm{cw}}
    = \fprod{\grad{\cL}{T} R_\textrm{cw}^\top, \partial J},
\quad
\textrm{where}~ 
\grad{\cL}{T}
= \grad{\cL}{\Sigma'} T\Sigma^\top + \grad{\cL}{\Sigma'}^\top T \Sigma.
\end{equation}

We continue propagating through $J$ for camera coordinates $t \in \mathbb{R}^4$
for the contribution through $\Sigma'$ to the gradients of $t$:
\begin{align*}
\grad{J}{t_x}
=
    \begin{bmatrix}
        0 & 0 & -f_x / t_z^2 \\
        0 & 0 & 0
    \end{bmatrix}, \quad
\grad{J}{t_y}
=
    \begin{bmatrix}
        0 & 0 & 0 \\
        0 & 0 & -f_y / t_z^2
    \end{bmatrix}, \quad
\grad{J}{t_z}
=
    \begin{bmatrix}
        -f_x / t_z^2 & 0 & 2 f_x t_x / t_z^3 \\
        0 & -f_y / t_z^2 & 2 f_y t_y / t_z^3
    \end{bmatrix},
\quad
\grad{J}{t_w} = \mathbf{0}^{2\times 3}.
\end{align*}

We can now sum the two gradients $\grad{\cL_{\mu'}}{t}$ and $\grad{\cL_{\Sigma'}}{t}$ into $G = \grad{\cL}{t}$,
and compute the full gradients with respect to the 3D mean $\mu$ and the view matrix $T_\textrm{cw}$.
We have that $t = T_\textrm{cw} q$, where $q = \begin{bmatrix} \mu&1 \end{bmatrix}^\top$.
\begin{align}
    \partial \cL &= \fprod{G, \partial t}  = \fprod{G, \partial (T_\textrm{cw} q)}\\
    &= \fprod{G q^\top, \partial T_\textrm{cw}} + \fprod{T_\textrm{cw}^\top G, \partial q}.
\end{align}

The gradients with respect to $T_{\textrm{cw}}$ and $\mu$ are then
\begin{equation}
\grad{\cL}{T_\textrm{cw}} = \grad{\cL}{t} q^\top \in \mathbb{R}^{4\times 4},
\quad
\grad{\cL}{\mu} = R_\textrm{cw}^\top \begin{bmatrix}
    \grad{\cL}{t_x}&
    \grad{\cL}{t_y}&
    \grad{\cL}{t_z}
\end{bmatrix}^\top \in \mathbb{R}^3.
\end{equation}

\subsubsection{Scale and rotation gradients}
Now we have $\Sigma = M M^\top$ and $\grad{\cL}{\Sigma}$.
Letting $G = \grad{\cL}{\Sigma}$, we have
\begin{align*}
    \partial \cL &= \fprod{G, \partial \Sigma} \\
    &= \fprod{G, (\partial M) M^\top + M (\partial M^\top)} \\
    &= \fprod{G M + G^\top M, \partial M}
\end{align*}
which gives us
\begin{equation}
\grad{\cL}{M} = \grad{\cL}{\Sigma} M + \grad{\cL}{\Sigma}^\top M.
\end{equation}

Now we have $M = RS$, with $G = \grad{\cL}{M}$ as
\begin{align*}
    \partial \cL &= \fprod{G, \partial M} \\
    &= \fprod{G, (\partial R) S} + \fprod{G, R (\partial S)} \\
    &= \fprod{G S^\top, \partial R} + \fprod{R^\top G, \partial S}
\end{align*}
which gives us
\begin{equation}
\grad{\cL}{R} = \grad{L}{M} S^\top, \quad
\grad{\cL}{S} = R^\top \grad{L}{M}.
\end{equation}

The Jacobians of the rotation matrix $R$ wrt the quaternion parameters $q = (w, x, y, z)$ are
\begin{align*}
    \grad{R}{w}
    = 2 \begin{bmatrix}
        0 & -z & y \\
        z & 0 & -x \\
        -y & x & 0
    \end{bmatrix}, \quad
    \grad{R}{x}
    = 2 \begin{bmatrix}
        0 & y & z \\
        y & -2x & -w \\
        z & w & -2x
    \end{bmatrix}, \quad
    \grad{R}{y}
    = 2 \begin{bmatrix}
        -2y & x & w \\
        x & 0 & z \\
        -w & z & -2y
    \end{bmatrix}, \quad
    \grad{R}{z}
    = 2 \begin{bmatrix}
        -2z & -w & x \\
        w & -2z & y \\
        x & y & 0
    \end{bmatrix}.
\end{align*}

The Jacobians of the scale matrix $S$ with respect to the scale parameters $s = (s_x, s_y, s_z)$ are
\begin{equation}
\grad{S}{s_j} = \delta_{ij}, 
\end{equation}
which select the corresponding diagonal element of $\grad{\cL}{S}$.

\section{Conclusion}
This report has provided the detailed derivations used in implementing differentiable Gaussian splatting in the \texttt{gsplat} library.
We have covered the essential aspects for both forward and backward passes, offering a comprehensive reference for researchers and practitioners.
Furthermore, we have provided a complete Python API exposing all computational components of forward and backward rasterization.
This API is designed to facilitate modification of each part of the computational graph to encourage further development and improvement.

\section{Acknowledgements}
This project was funded in part by NSF:CNS-2235013. We thank Brent Yi for revisions and comments. Thank you to \texttt{nerfstudio} team members Matias Turkulainen and Zhuoyang Pan for validation and integration tests, as well as Justin Kerr and Ruilong Li for integrating into the \texttt{nerfstudio} platform.

\bibliographystyle{plainnat}
\bibliography{ref}

\end{document}